\documentclass[twocolumn,prd,nofootinbib,showpacs]{revtex4}
\usepackage{amsmath}

\newcommand{\vhat}[1]{\hat{{#1}}}

\usepackage{graphicx}

\newcommand{\beq}{\begin{equation}}
\newcommand{\eeq}{\end{equation}}
\newcommand{\beqa}{\begin{eqnarray}}
\newcommand{\eeqa}{\end{eqnarray}}
\newcommand{\lwmap}{\lambda_{\mbox{\scriptsize WMAP}}}

\begin{document}
\title{Cosmic microwave background constraints on cosmological models
\\ with large-scale isotropy breaking}
\author{Haoxuan Zheng}
\author{Emory F. Bunn}
\email{ebunn@richmond.edu}
\affiliation{Physics Department, University of Richmond, Richmond, VA  23173}

\begin{abstract}
Several anomalies appear to be present in the large-angle cosmic
microwave background (CMB) anisotropy maps of WMAP, including
the alignment of large-scale multipoles.
Models in which isotropy is spontaneously broken (e.g., by a 
scalar field) have been proposed as explanations for these
anomalies, as have models in which a preferred
direction is imposed during inflation.  
We examine models inspired by these, in which isotropy is broken
by a multiplicative factor with dipole and/or quadrupole terms.  We evaluate
the evidence provided by the multipole alignment
using a Bayesian framework, finding that the 
evidence in favor of the model is generally weak.  
We also compute approximate changes in estimated cosmological parameters
in the broken-isotropy models.  Only the overall normalization
of the power spectrum is modified significantly.
\end{abstract}
\pacs{98.80.-k,
98.70.Vc, 
98.80.Es, 
95.85.Bh 
}
\maketitle

\section{Introduction}

Our understanding of cosmology has advanced extremely rapidly
in the past decade.  These advances are due in large part to
observations of cosmic microwave background (CMB) anisotropy,
particularly the data from the Wilkinson Microwave Anisotropy
Probe (WMAP)
\cite{wmap1yr1,wmap1yr2,wmap1,wmap7yrbasic}.
As a result of these and other observations, a
``standard model'' of cosmology has emerged, consisting of a Universe dominated
by dark energy and cold dark matter, with a nearly scale-invariant
spectrum of Gaussian adiabatic perturbations
\cite{wmap7yrparams,wmap7yrinterp} of the sort
that would naturally be produced in an inflationary epoch.

The overall consistency of the CMB data with this model is quite
remarkable. In particular, the CMB observations are very nearly
Gaussian, and the angular power spectrum matches theoretical
models very well from scales of tens of degrees down to arcminutes.
However, several anomalies have been noted on the largest
angular scales, including a lack of large-scale power
\cite{wmap1yr2,copi2,dOCTZH}, alignment of low-order multipoles
\cite{copi1,schwarz,dOCTZH,hajian}, and hemispheric asymmetries
\cite{eriksen2004,freeman,hansen}.
Some anomalies seem to be associated with the ecliptic plane, suggesting
the possibility of a systematic error associated with the WMAP
scan pattern, perhaps related to coupling of the scan pattern
with the asymmetric beam \cite{hanson}.  If the anomalies
have cosmological significance, then naturally the correlation
with the ecliptic plane must be a coincidence.

The significance of and explanations for these puzzles
are hotly debated.  In particular, it is difficult to know how to interpret
a posteriori statistical significances: when a statistic is invented
to quantify an anomaly that has already been noticed, the low $p$-values
for that statistic cannot be taken at face value.

One can (and from a formal statistical point of view, arguably
one must) dismiss this entire subject 
on the ground that all such anomalies are characterized only by
invalid a posteriori statistics \cite{wmap7yranomalies}.
Nonetheless, the number and nature of the anomalies (in particular, the
fact that several seem to pick out the same directions on the sky)
seem to suggest that there may be something to explain in the data.
Given the potential importance of new discoveries about the Universe's
largest observable scales, and the difficulty in obtaining a new data
set that would allow for a priori statistical analysis, we believe
that the potential anomalies are worth further examination.
In this paper, we will tentatively assume that there is a need for an 
explanation and consider what that explanation might be.

One of the most robust of the large-scale anomalies found in WMAP
is a lack of large-scale power, as quantified either by the low
quadrupole or the vanishing of the two-point correlation function at large
angles \cite{wmap1yr2,copi2,dOCTZH}.  If this anomaly is real, then
it provides strong evidence {\it against} a broad class of nonstandard
models.  To be specific, all models in which a statistically independent
contaminant (whether due to a foreground, systematic error, or exotic
cosmology) is added to the data will necessarily fare worse than
the standard model in explaining this anomaly \cite{gordon,bunnbourdon}.
There is a simple reason for this: a statistically independent
additive contaminant always increases the root-mean-square power
in any given mode, reducing the probability of finding low power.

It is natural, therefore, to seek an explanation of the anomalies
among models that do not involve a mere additive contaminant.  One
simple phenomenological model is a multiplicative contaminant, in
which the original statistically isotropic CMB signal $T^{(0)}(\theta,\phi)$ is
modulated by a multiplicative factor, leading to an observed signal
\beq
T(\theta,\phi)=f(\theta,\phi)T^{(0)}(\theta,\phi).
\label{eq:modulate}
\eeq

This model arises naturally in the framework of spontaneous isotropy breaking
by a scalar field \cite{gordon}.  
Moreover, models based on the 
existence of a vector field specifying a preferred direction
during inflation \cite{ackerman,bohmer} produce similar modulation, but with
$f$ having specifically a quadrupolar form.   To be precise, the modulation
in these models takes place in the primordial power spectrum $P(\vec k)$,
which acquires a quadrupolar dependence on the direction of $\vec k$.
The full effect on the CMB anisotropy is more complicated than the
above model, but the dominant effect on large scales is, at least
approximately, a quadrupolar moduloation of the above form.\footnote{The
stability of the specific
model of ref.\ \cite{ackerman} has been questioned \cite{acwunstable}; 
nonetheless, we believe it is worthwhile to consider models of this
general class.}

Since our goal is to explain the observed large-scale anomalies while
maintaining the success of the standard model on smaller scales, it is
natural to consider models in which $f$ has
power only on large scales.  We will consider three classes of model: one
in which $f$ has only monopole and dipole terms, one in which it has
monopole and quadrupole, and one in which it has 
all three.  We will refer to these as the dipole-only, quadrupole-only,
and dipole-quadrupole models.
The quadrupole-only model is inspired by the theory
of a preferred direction during inflation, while the others are 
inspired by the general isotropy-breaking framework.

This paper addresses the following central question.  Do the
broken-isotropy models provide an explanation for one of the main
observed anomalies, namely the surprising 
alignment between the quadrupole and octupole (multipoles $l=2,3$)?
To examine this
question, we choose statistics to quantify the anomaly and use these
statistics to assess goodness of fit of the data to the different
models.  Several different statistics are chosen in order to assess the 
robustness of the results.

Because the statistics are most naturally computed in
spherical harmonic space, we use the all-sky internal linear
combination (ILC) maps from the five-year WMAP data release
\cite{wmap5yrbasic}.  There is bound to be residual foreground
contamination in the ILC maps \cite{eriksenilc04,eriksenilc05}.  
Section \ref{sec:foregrounds} contains a
brief discussion of the effects of this contamination.

Naturally,
because the anisotropic models have more free parameters than
the standard model (and indeed include the standard model
as a special case), there will 
generically be parameter choices that make the anisotropic
model fit the data better.  We adopt the Bayesian evidence criterion
to assess whether this improved fit is sufficient to justify the
additional complexity of the anisotropic model.  Bayesian evidence has been
used in addressing this sort of question in the past 
\cite{landmag,eriksen2007,hoftuft}.
Although some controversy has arisen over its use in cosmology (e.g., 
\cite{marshall,lidmukpar,liddle,efstathioubayes}), in this context it is both a
simple and a natural criterion to adopt.

In some
cases, the Bayesian evidence ratios are greater than one, meaning
that one's assessment of the probability that the broken-isotropy
models are true should rise as a result of the CMB anomalies.  However,
in all cases, the improvement is modest, 
providing
at most weak support for the adoption of the anisotropic models.  

We also consider the changes in parameter estimates that 
would arise if the anisotropic models are correct.  To be specific,
because we assume that the modulation is a perturbation
to the standard model, 
we assume that the unmodulated temperature map $T^{(0)}$ is derived
from the cosmological parameters in the usual way -- i.e., its
power spectrum is given by CMBFAST \cite{cmbfast}.  If
there is a nonconstant modulation function $f$, then parameter estimates
from based on the observed data will naturally differ from the true
values.  We estimate the resulting parameter shifts, finding
them to be minor.

The remainder of this paper is structured as follows.  Section 
\ref{sec:simulate} specifies precisely the anisotropic models
under consideration and describes how we simulate these models.  In Section
\ref{sec:bayes}, we review the method for computing Bayesian evidence
ratios.  Section \ref{sec:align} contains our 
main results, indicating the degree to which the multipole alignment, quantified in several different ways, favor the broken-isotropy models.
In Section \ref{sec:params} we quantify the degree to which 
best-fit cosmological parameters are modified by changing from
the standard model to the broken-isotropy models.  Section \ref{sec:foregrounds}
discusses some aspects of the issue of foreground contamination.  Finally, we
provide a brief discussion of our results in Section \ref{sec:discussion}.

\section{Simulating anisotropic models}
\label{sec:simulate}

The statistical properties of a CMB map
are most easily expressed in terms of the spherical harmonic
expansion,
\beq
T(\theta,\phi)=\sum_{l=0}^\infty\sum_{m=-l}^l
a_{lm}Y_{lm}(\theta,\phi).
\label{eq:texpansion}
\eeq
The monopole ($l=0$) term in the sum is simply the average temperature
over the sky, and the dipole ($l=1$) terms cannot be separated
from the kinematic dipole due to our motion with respect to the CMB
``rest'' frame.  These terms are typically removed from the data, so that
in practice the sum starts at $l=2$.  For compactness, we will
generally abbreviate such double sums as $\sum_{l,m}$, not writing 
the limits explicitly unless
confusion may arise.

In the standard model, the CMB map $T^{(0)}(\theta,\phi)$ 
is a realization of a statistically
isotropic Gaussian random process.  This means that the spherical
harmonic coefficients
$a^{(0)}_{lm}$ of this map 
are independent Gaussian random variables with mean zero
and variances that depend only on $l$:
\beq
\langle |a^{(0)}_{lm}|^2\rangle = C_l^{(0)},
\eeq
where $\langle\cdot\rangle$ denotes an ensemble average and $C_l^{(0)}$ is
the power spectrum.  

In broken-isotropy models, on the other hand, we assume that the observed
field is related to the above statistically isotropic
expression according to equation (\ref{eq:modulate}).
We expand the modulation function in spherical harmonics,
\beq
f(\theta,\phi) = 1+\sum_{l,m} f_{lm}Y_{lm}(\theta,\phi).
\label{eq:fexpansion}
\eeq
We assume that the modulation function is normalized to have mean one,
so that the above sum starts at $l=1$.  (Equivalently, we could omit the 
$1+$ in the above expression and start the sum at $l=0$ with $f_{00}=
\sqrt{4\pi}$.)
We will assume that the coefficients $f_{lm}$ are independent Gaussian random
variables with a power spectrum
\beq
C_l^{(f)}\equiv\langle |f_{lm}|^2\rangle.
\eeq
As noted in the Introduction, we consider models in which $f$ 
has only dipole and/or quadrupole terms.
We parameterize these terms with
parameters $\sigma_1,\sigma_2$, giving the rms values of $f_{lm}$ relative
to a scale-invariant spectrum $C_l^{(f)}\propto[l(l+1)]^{-1}$:
\beq
\sigma_1^2=2C_1^{(f)},\qquad\qquad
\sigma_2^2=6C_1^{(f)}.
\eeq
Because the spherical harmonics have root-mean-square (rms)
value $(4\pi)^{-1/2}$, these modulations have rms amplitudes $(8\pi)^{-1/2}
\sigma_1=0.20\sigma_1$ and $(24\pi)^{-1/2}\sigma_2=0.12\sigma_2$
respectively.

The spherical harmonic coefficients of the observed sky
are found as usual by spherical harmonic orthonormality:
\begin{eqnarray}
a_{lm}&=&\int d\Omega\,T(\theta,\phi)Y_{lm}^*(\theta,\phi)
\\&=&\int d\Omega\,f(\theta,\phi)T^{(0)}(\theta,\phi)Y_{lm}^*(\theta,\phi).
\end{eqnarray}
Expanding the functions $T^{(0)}$ and $f$ in spherical harmonics,
we find that
\beq
a_{lm}=\sum_{l_1,m_1}\sum_{l_2,m_2}a_{l_1m_1}^{(0)}f_{l_2m_2}
I_{l_1m_1l_2m_2lm},
\label{eq:almyyy}
\eeq
where $I_{l_1m_1l_2m_2lm}$ represents
an integral over three spherical harmonics, which ca
be expressed in terms of Wigner 3-$j$ symbols \cite{zare}:
\begin{eqnarray}
I_{l_1m_1l_2m_2lm}&\equiv&
\int
Y_{l_1m_1}Y_{l_2m_2}Y_{lm}^*\,d\Omega\\&=&
\left[{\left(2l+1\right)\left(2l_1+1\right)\left(2l_2+1\right)\over
4\pi}\right]^{1\over 2}\times\nonumber\\
&&\quad\begin{pmatrix} l&l_1&l_2\\0&0&0
\end{pmatrix}\begin{pmatrix}
l&l_1&l_2\\-m&m_1&m_2
\end{pmatrix}
\label{eq:intyyy}
\end{eqnarray}

The quadruple sum in equation (\ref{eq:almyyy}) has very
few nonzero terms and hence can be quickly evaluated.
Because our model includes only low-$l$ power in $f$,
the sum over $l_2$
ranges from 0 to at most 2.  Moreover, the Wigner 3-$j$ symbols
vanish unless certain
conditions are satisfied.  First, 
$(l,l_1,l_2)$ must satisfy a triangle inequality, so
that the sum over $l_1$ ranges from $l-2$ (or zero, whichever is greater)
to $l+2$. Second, the first of the two 3-$j$ symbols vanishes
unless $l+l_1+l_2$ is even.  
Finally, the constraint $m=m_1+m_2$ must be satisfied.

\section{Bayesian evidence}
\label{sec:bayes}

Our goal in this paper will be to compare the standard model (the
null hypothesis) with the class of broken-isotropy models.  Naturally,
because the latter class is broader, and indeed includes the null
hypothesis as a limiting case, there will generically be members
of the class that fit the data better than the standard model.
The Bayesian evidence provides a framework for assessing 
whether the better fit found in the more complicated
model is worth the Occam's-razor ``cost.''  We now briefly
review this approach to model comparison.

Suppose that we have a model $M$ that depends on a set of
parameters $\boldsymbol{\theta}$.  Given a data set $D$, we define
the evidence of the model to be the probability density
of $D$, given the model $M$:
\beq
E(M) = \int d\boldsymbol{\theta}\,P(D|M,\boldsymbol{\theta})
\pi(\boldsymbol{\theta}|M).
\eeq
In this expression, $P$ is the likelihood function -- that is, the
probability density for the data given the choice of model and parameters
-- and $\pi$ is the prior probability density of the model parameters.
It may be helpful to keep track of dimensions in 
these expressions.  The prior $\pi$ has dimensions of probability
per unit volume in \textit{parameter space}, while $P$ and $E$ have
dimensions of probability per unit volume in \textit{data space}.

Bayes's theorem says that the posterior probability of the model
is proportional to the product of the model's prior probability and
the evidence.  Suppose now that we have two models $M_1,M_2$ in mind,
and imagine that,
before looking at the data set $D$, we regarded these models as equally
probable.  Then the evidence ratio 
\beq
\Lambda\equiv{E(M_1)\over E(M_2)}
\eeq 
is equal to 
the ratio of posterior probabilities.  

In the case we consider in this paper, the two models are the standard
model and the broken-isotropy model.  The reader (like the writers)
probably does
not assign equal prior probabilities to these two models:
in the absence of the WMAP anomalies, most of us probably thought
that the broken-isotropy model was less likely.  Even in this case,
the evidence ratio still tells us 
by what factor
the broken-isotropy model goes up in our estimation (relative to the
isotropic model) as a result
of the data.

The Bayesian evidence automatically accounts for the degree of 
complexity of the model, in the sense that models with a large
parameter space will be automatically downweighted compared
to those with a small parameter space.  To see this heuristically,
suppose that the prior probability $\pi$ is approximately
flat over some volume $V_p$ in parameter space that
is much larger than the range over which the likelihood
function is large.
Then since the probability
distribution is normalized,
\beq
\int d\boldsymbol{\theta}\,\pi(\boldsymbol{\theta}|M)=1,
\eeq
we can estimate $\pi\sim V_p^{-1}$ over the range where the
integrand is significant.  Thus we can crudely estimate
\beq
E(M)\sim V_p^{-1}\int d\boldsymbol{\theta}\,P(D|M,\boldsymbol{\theta})
\sim {P_{\rm max}V_L\over V_p},
\eeq
where $P_{\rm max}$ is the peak of the likelihood function and
$V_L$ is an estimate of the volume in parameter space over which
the likelihood differs significantly from zero.  If we consider
two models with similarly good fits to the data (i.e., similar
values of $P_{\rm max}$), the one with a higher value of the
ratio $V_p/V_L$ will have the higher value of the evidence.
In other words, the 
Bayesian evidence disfavors 
models with a large volume of ``wasted'' parameter
space.  When comparing models with parameter spaces of different
dimensions, the one with a higher-dimensional parameter space will
typically be disfavored, unless it provides 
a much better fit to the data (i.e., has a large $P_{\rm max}$) or 
it provides a reasonably good fit to the data
over most of the parameter space.

Below, we will use Bayesian evidence ratios to assess whether
the multipole alignment anomaly significantly
favors the adoption of the more complicated broken-isotropy models,
using the following procedure.  We 
define a statistic $s$ that describes the anomaly.
Since the null (statistically isotropic) hypothesis $M_0$ has
no free parameters, the
evidence for it
is simply the probability density of the statistic under
that hypothesis:
\beq
E_0=P(s|M_0).
\eeq
For some
choices of statistic, this probability
density can be computed analytically, but in general it must be
estimated from simulations.

\begin{figure*}[t]
\begin{tabular}{cc}
(a) & (b) \\
\includegraphics[width=3in]{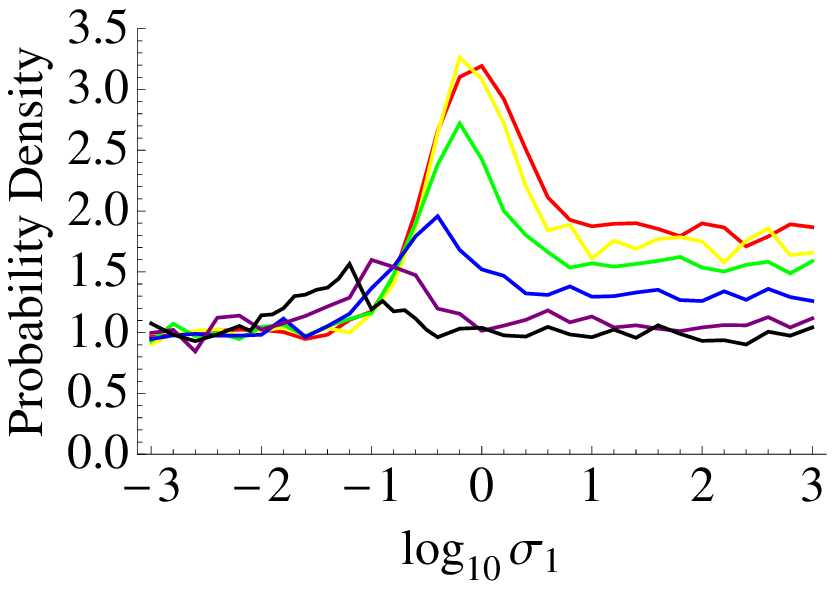}&
\includegraphics[width=3in]{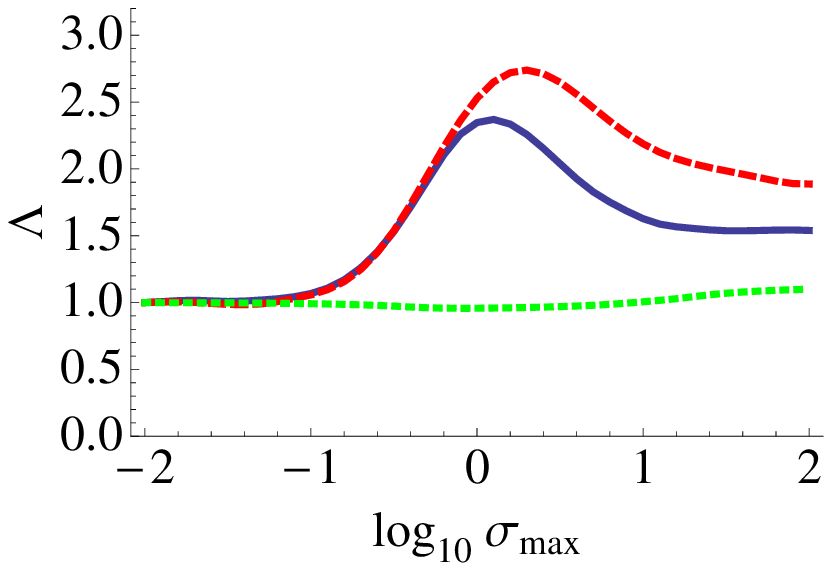}
\end{tabular}
\caption{(a) Values of the probability density
function (pdf) for the multipole alignment
statistic $\lambda$, evaluated in an interval of 
width $\delta\lambda=0.02$ about $\lambda=0.985$, from
$10^4$ simulations for each choice of parameter values.  
The different curves
correspond to $\sigma_2=(0,e^{-1},1,e,e^2,e^3)=(0,0.37,1,2.7,.7.4,20)$ 
(from highest to lowest
pdf at the right of the plot).  (b)The Bayesian
evidence for the anisotropic models.  The solid curve is for
the dipole-only model, the dashed curve is for the dipole-quadrupole
model, and the dotted curve is for the quadrupole-only model.}
\label{fig:momentum}
\end{figure*}

We now consider the evidence $E_1\equiv E(M_1)$
for the broken-isotropy model.  
Let us first examine the models in which $f$
has only power in one multipole (i.e., the dipole-only and quadrupole-only
models).  The parameter space $\boldsymbol{\theta}$
for this model consists of the single parameter $\sigma_j$, where $j=1$
for the dipole-only model and 2 for the quadrupole-only model.  
To compute the evidence for this model, we must choose a prior
$\pi(\sigma_j)$.  We adopt a uniform prior on some range
$\sigma_j\in [0,\sigma_{\rm max}]$:
\beq
\pi(\sigma_j)=\begin{cases}\sigma_{\rm max}^{-1} & 0<\sigma_j<\sigma_{\rm max},
\\
0 & \mbox{otherwise.}
\end{cases}
\label{eq:sigprior1}
\eeq
For the dipole-quadrupole model, we follow
a similar procedure, adopting a prior on $\boldsymbol{\theta}=(\sigma_1,
\sigma_2)$ of
\beq
\pi(\sigma_1,\sigma_2)=\begin{cases}\sigma_{\rm max}^{-2} & 
0<\sigma_1,\sigma_2<\sigma_{\rm max},\\
0 & \mbox{otherwise.}
\end{cases}
\label{eq:sigprior2}
\eeq

Since it is not obvious what cutoff $\sigma_{\rm max}$ to choose,
we plot the evidence ratio $E_1/E_0$
as a function of this parameter.  We
regard the maximum value of the evidence ratio as an upper bound
on the true evidence ratio.  From the heuristic argument
above we expect the evidence ratio to decline for very
large values of $\sigma_{\rm max}$, since these models
presumably have ``wasted'' parameter space.

\section{Results}
\label{sec:align}

Various statistics have been used in the past to characterize the 
observed alignment of the $l=2$ and $l=3$ multipoles in the WMAP data.
We focus on two categories of statistic: one based on finding
the directions that maximize the angular momentum \cite{dOCTZH} for each
multipole (Section \ref{subsec:angmom}), and one based on multipole vectors \cite{schwarz,copi1} (Section \ref{subsec:mpv}).  

\subsection{Angular momentum}
\label{subsec:angmom}
For any given
multipole $l$, consider the map obtained by keeping just
the corresonding coefficients in the spherical harmonic expansion,
\beq
T_l(\theta,\phi)=\sum_{m=-l}^la_{lm}Y_{lm}(\theta,\phi).
\eeq
The maps $T_2$ and $T_3$ are each observed to have fluctuations that lie
predominantly in a single plane, and moreover the planes associated
with these two multipoles seem to be aligned 
\cite{schwarz,copi1,dOCTZH,hajian}.  The idea
of the maximum-angular-momentum statistic is
to quantify that alignment
by defining for each $l$
an axis perpendicular to the plane picked put by the map $T_l$.

Consider a particular map $T_l$.  
For any given direction, specified by a unit vector $\vhat n$, 
we imagine rotating the 
map to bring $\vhat n$ to the $z$ axis.  Let
$a_{lm}^R$ represent
the spherical harmonics in the rotated coordinate system, which
can be efficiently computed by applying an appropriate
Wigner $D$ matrix to the unrotated $a_{lm}$'s \cite{zare}.
We compute the 
the ``angular momentum'' of the rotated map about the $z$ axis:
\beq
L_z^2(\vhat n)=\sum_{m=-l}^l m^2 |a_{lm}^R|^2.
\eeq
The direction $\vhat n$ that maximizes $L_z^2$ is taken to be the 
axis $\vhat n_l$ for the given multipole.  Note that because $L^2_z(\vhat n)
=L^2_z(-\vhat n)$, the vector $\vhat n_l$ is only defined
up to an overall sign.  

We use the statistic $\lambda = |\vhat n_2\cdot \vhat n_3|$ 
to assess the degree to which the fluctuations in the quadrupole
and octupole are aligned.
In any statistically isotropic model, we expect the directions $\hat n_l$
to be indepent and uniformly distributed over the unit sphere,
which
implies that $\lambda$ is uniformly distributed
on the interval $[0,1]$.  The value in the actual WMAP data is
surprisingly large at $\lwmap= 0.985$.  

For any given choice of parameters $(\sigma_1,\sigma_2)$, we can 
simulate a large number of maps and determine
the probability density function (pdf) 
of the statistic $\lambda$.  Specifically, we can estimate the
average pdf in an interval of with $\delta\lambda$ around the value
$\lwmap$ simply by finding the fraction of all simulations
yielding values in the range $[\lwmap-{1\over 2}\delta\lambda,
\lwmap+{1\over 2}\delta\lambda]$.
Figure \ref{fig:momentum}a
shows the resulting pdfs, based on $10^4$ simulations for each point in
parameter space, with $\delta\lambda
=0.02$. The Poisson noise in this estimation
process is visible as $\sim 7\%$ scatter in the points in this plot.  
Histograms of the simulation results confirm that the pdfs are smooth over
scales much larger than $\delta\lambda$, so interpreting the average pdf
as the pdf at the given point is reasonable.

\begin{figure*}[t]
\begin{tabular}{cc}
(a)&(b)\\
\includegraphics[width=3in]{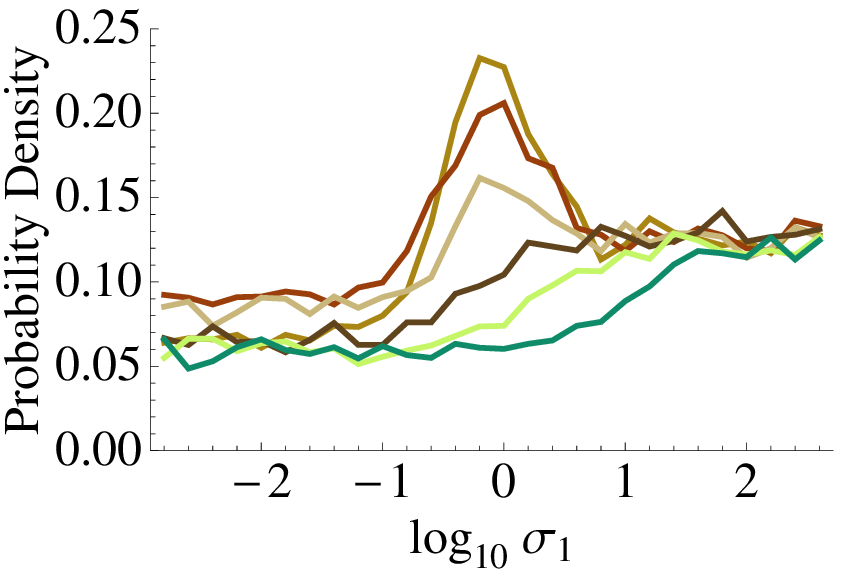}&
\includegraphics[width=3in]{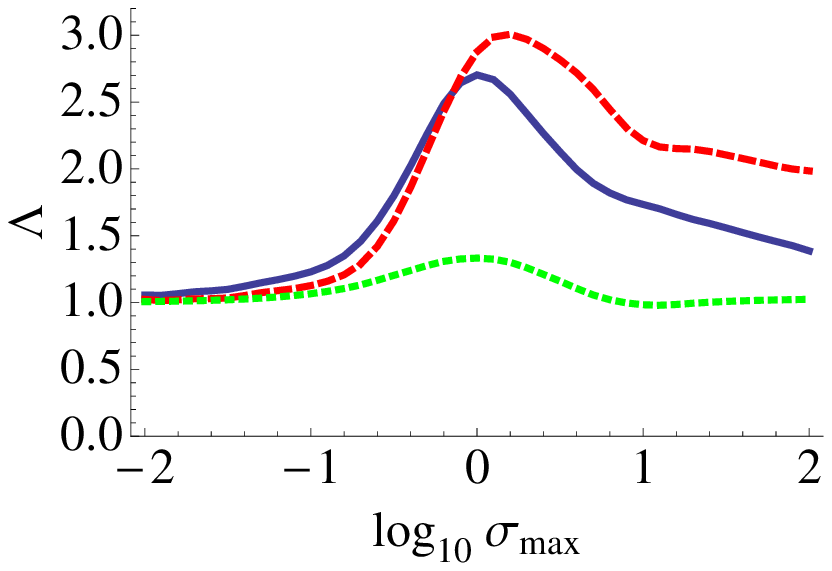}
\end{tabular}
\caption{
(a)
Probability density for the Schwarz et al. \cite{schwarz} multipole vector
statistic $S$, evaluated at the value found in the WMAP data.  From top to 
to bottom at $\log_{10}\sigma_1=0$, the curves correspond to $\sigma_2=(0,
0.25,0.50,1.,2.0,4.0,16)$.
(b) Bayesian evidence for 
anisotropic models.  Results are plotted for the dipole-quadrupole (solid),
dipole-only (dashed), and quadrupole-only (dotted) models.}
\label{fig:schwarz}
\end{figure*}

Since the pdf under the
null hypothesis is equal to 1, this quantity can be interpreted
as a Bayesian evidence ratio comparing the model with the given
values of $(\sigma_1,\sigma_2)$ to the null hypothesis.  

As Figure \ref{fig:momentum}a shows, for some 
choices of parameter, the evidence ratio exceeds $3$.  However, this
overstates the evidence in favor of the broken-isotropy model.  As described
in Section \ref{sec:bayes}, the correct procedure is to treat
$\sigma_1,\sigma_2$ as unknown parameters with a given prior
distribution, and integrate over that prior to get the 
evidence.  The integration is performed numerically, after interpolating
between the likelihood 
estimates found for the various values of $(\sigma_1,\sigma_2)$.

Figure \ref{fig:momentum}b shows the result of this calculation.
The quantity on the horizontal axis is the prior
cutoff $\sigma_{\rm max}$ of equation (\ref{eq:sigprior1}) or
(\ref{eq:sigprior2}).  Because each Bayesian evidence ratio is an integral
over the likelihood function, the effect of Poisson noise due to the
finite number of simulations is greatly reduced.

In the dipole-quadrupole case 
(where both $\sigma_1,\sigma_2$ are free parameters),
the
Bayesian evidence ratio has a maximum value of  $\sim 2.4$
at $\sigma_{\rm max}\sim 1$.  (Recall that, as noted
in Section \ref{sec:simulate}, $\sigma=1$ corresponds
to only 10-20\% modulation.) 
The dipole-only model (in which only $\sigma_1$
varies)
fares a bit better, with evidence ratio peaking at $\sim 2.7$.
Even if we take this maximum value
as the true evidence ratio, it is still only modest support for
the broken-isotropy model.  The quadrupole-only model
shows
no significant improvement at all over the standard model (as we
could have predicted
from Figure \ref{fig:momentum}a, in which all curves approach
the standard-model value of 1 for low $\sigma_1$).

\subsection{Multipole vectors}
\label{subsec:mpv}

To test the robustness of this result, we can use a different
approach to quantify the multipole alignment.  For each
multipole $l$, the map $T_l$ can be used to define $l$ unit vectors,
generally
called ``multipole vectors'' \cite{copi1}. The multipole vectors for
each $l$ can be used to characterize the orientation of that multipole,
and thus to characterize the quadrupole-octupole alignment.

There are multiple different ways of using the multipole
vectors to define an alignment statistic.  The original work on 
the subject \cite{copi1} used an elaborate procedure involving the
assessment of
several different combinations of
dot and cross products of the multipole vectors.  Subsequent
work by members of the same group \cite{schwarz} focused on a smaller
subset of these possibilities.  We have chosen to implement the ``robust
and more conservative'' statistic used in the latter work.  We now describe
this statistic.

Let $\vhat{v}^{(l,j)}$ ($1\le j\le l$) represent the $j$th multipole vector
for multipole $l$.  For any given $l$, we consider all $l(l-1)/2$ distinct
cross products of multipole vectors $\vec w^{(l,i,j)}\equiv
\hat v^{(l,i)}\times\hat v^{(l,j)}$ ($1\le i<j\le l$).  Alignment of the
quadrupole and octupole planes can be characterized by the absolute values
of the dot product of the one quadrupole cross product, $\hat w^{(2,1,2)}$, with
each of the three octupole cross products $\hat w^{(3,i,j)}$.  (The 
absolute value is necessary because the multipole vectors, and
hence the cross products, are specified only up to an overall sign.)
Following ref.\ \cite{schwarz}, we therefore define a statistic
\beq
S=|\vec w^{(2,1,2)}\cdot\vec w^{(3,1,2)}|+
|\vec w^{(2,1,2)}\cdot\vec w^{(3,1,3)}|+
|\vec w^{(2,1,2)}\cdot\vec w^{(3,2,3)}|
\label{eq:schwarzstat}
\eeq
The value of this statistic for the WMAP data is $S_{\mbox{\scriptsize WMAP}}=
2.233$.  Based on Monte Carlo simulations, we find this to be inconsistent
with the standard isotropic model at $99.3\%$ confidence.  These
values differ only slightly from those found in ref.~\cite{schwarz}
($S_{\mbox{\scriptsize WMAP}}=2.396$, ruled out at $99.87\%$ confidence),
which used an earlier data release and a different foreground
removal process \cite{tegmarkmap}.

\begin{figure*}[t]
\begin{tabular}{cc}
(a)&(b)\\
\includegraphics[width=3in]{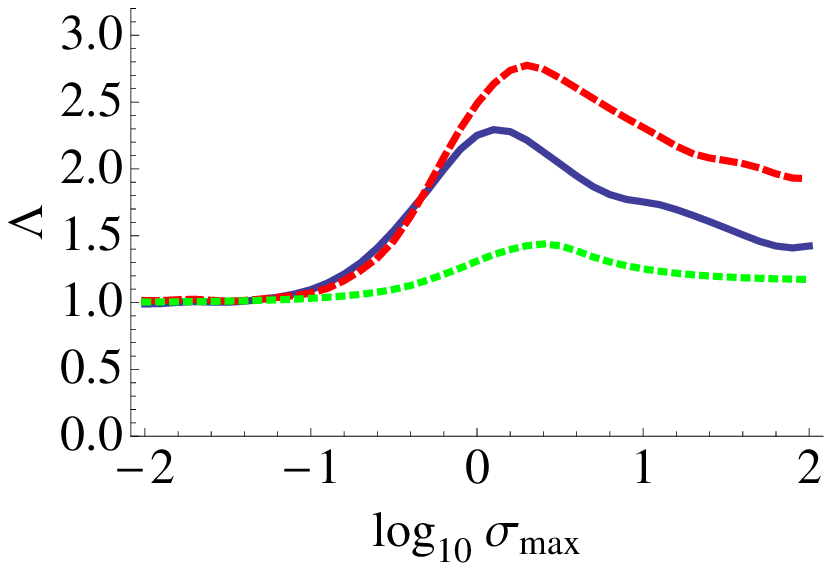}&
\includegraphics[width=3in]{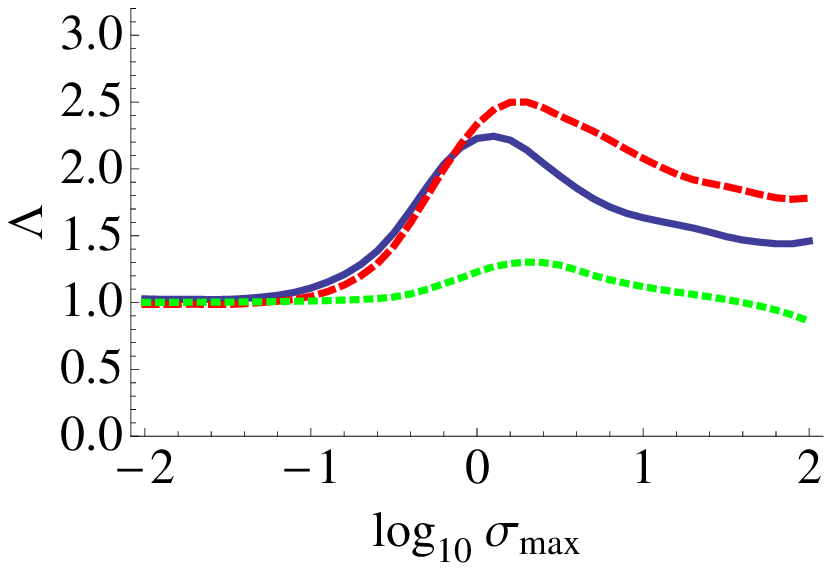}
\end{tabular}
\caption{Bayesian evidence ratios calculated using multipole
vectors.  In (a), the joint probability density
for $(A,p)$, the alignment and planarity statistics was used.  In 
(b), only the statistic $A$ was used.
Results are 
plotted for the dipole-quadrupole (solid), dipole-only (dashed),
and quadrupole-only (dotted) models.}
\label{fig:mpvbayes}
\end{figure*}

Figure \ref{fig:schwarz} shows the result of Bayesian evidence calculations
based on this statistic.  At 
each point $(\sigma_1,\sigma_2)$ in parameter space, 
the pdf (i.e., the likelihood function) was
evaluated from
$10^4$ simulations, by counting the number of times the statistic
was found in an interval of width $\delta S=0.15$ about the value
found in the WMAP data.
The results are qualitatively consistent
with those based on the angular momentum statistic, although with
slightly higher evidence ratios (i.e., slightly more favorable to 
the broken-isotropy models).  

There is of course some arbitrariness in the choice of the statistic $S$. 
In addition to $S$,
we devised an alternative set of statistics based on
the multipole vectors. The results based on these statistics
can be viewed as a test of the robustness of the results above.

In defining our statistics we were guided by a desire to characterize
the observed dipole-quadrupole alignment and the fact that the octupole
has been characterized as unusually planar.  Since we were
guided by these already-observed facts, of course, our choices are subject
to the same a-posteriori-statistics criticism as most other work in this
area.  We did, however, attempt to avoid exacerbating this problem with
further a posteriori choices: we devised our statistics blindly and used
only one statistic to characterize each of these two phenomena.

The two multipole vectors at $l=2$ define
a plane, and we let $\hat n_2$ be the unit vector perpendicular to that
plane.  To assess the multipole alignment, we need to define a similar
unit vector based on the three $l=3$ vectors.
We define $\hat n_3$ to be the unit vector that
is as nearly as possible perpendicular to these vectors by
minimizing the quantity 
\beq
p=\sum_{1\le i<j\le 3} (\hat n_3\cdot \hat v^{(3,i,j)})^2.
\eeq 

As in the angular momentum case, we define an alignment statistic to be
the absolute value of the dot product of these vectors:
\beq
A=|\hat n_2\cdot \hat n_3|.
\eeq
In addition, the statistic $p$ can be thought of as characterizing
the octupole planarity, with low values of $p$ corresponding to more
planar octupole patterns.

The value of $A$ for the real data is 0.97, which is somewhat anomalously
high since a uniform distribution on [0,1] is expected in the standard
model.  The statistic $p$, on the other hand, does not show anomalous
planarity: its value in the real data is 0.31, lying near the middle
of the distribution in simulations based on the standard model.

Since $p$ is quite consistent with the standard model, we would not
expect its inclusion in our analysis to improve the evidence for
any nonstandard models.  For completeness, we performed the Bayesian evidence
calculations using the joint probability density on $A$ and $p$ as
our input likelihood function, and also using the probablility densities
on $A$ and $p$ separately.  

The probability densities for each parameter
were calculated as with the previous statistics, by counting the
number of simulations yielding values in a small interval about the 
value in the true data.  In this case, we used $10^5$ simulations
at each data point, with $\delta A=\delta p=0.005$.  The joint probability
density was estimated as the product of the individual pdfs.  In principle,
the two statistics could be correlated, in which case this would
not be correct.  In practice, however, correlations were found
to be negligible for the models under consideration; in spot-checks
this approximation was found to be quite good.

Figure \ref{fig:mpvbayes} shows the result of Bayesian evidence computations
based on this statistic.  As expected, the results vary only slightly depending
on whether the planarity statistic is included.  In either case, the
strongest evidence ratio comes at $\sigma_{\rm max}\approx 1$
in the dipole-only model, but as before the
evidence ratios are modest, peaking at $\sim 2.7$ including both statistics
and $\sim 2.5$ using only the alignment statistic.
Results showing the planarity-only statistic are not shown but 
yield no significant enhancement in the evidence ratio.

The strong similarity in all of the evidence ratio plots suggests
that our results are insensitive to the precise way that the multipole 
alignment is characterized.

\section{Corrections to cosmological parameters in anisotropic models}
\label{sec:params}

In the anisotropic models under consideration, the power spectrum $C_l$
is modified by the modulation function $f$.  We assume that
the original, unmodulated power spectrum $C_l^{(0)}$, as opposed
to the measured power spectrum $C_l$, is produced
by the usual standard-model mechanism.  
In the anisotropic models, therefore,
the cosmological parameters estimated from the power spectrum will
differ from those in the standard model.  In this section we estimate
the changes in parameters as functions of $\sigma_1$ and $\sigma_2$.  
To be specific, we will assume that $C_l$ has been estimated
from the data and used to derive power spectrum estimates under
the standard assumptions of isotropy and Gaussianity.  We will
compute the corrections that must be applied to these parameter estimates
for nonzero $\sigma_1,\sigma_2$.
We find
that for reasonable values of $\sigma_1,\sigma_2$, all
parameters except the overall normalization undergo very small changes.

The changes in parameter values we compute depend on the assumption
that the modulation is the same across all angular scales.  If 
the modulation
exists only on large scales, with smaller scales
described by the unmodulated standard model,
then the changes in parameter estimates
will be even smaller than those found here.

To estimate the changes in parameter values, we assume that the
unmodulated CMB power spectrum is given by the standard model and can
be calculated from, e.g., CMBFAST \cite{cmbfast}.  We begin
by deriving the relationship between
the modulated (i.e., observed)
power spectrum $C_l$, the unmodulated power spectrum
$C_l^{(0)}$, and the power spectrum $C_l^{(f)}$ of the modulating
function.
We begin from equation (\ref{eq:almyyy}):
\beq
a_{lm}=\sum_{l_1,m_1}\sum_{l_2,m_2}a_{l_1m_1}^{(0)}f_{l_2m_2} 
I_{ll_1l_2mm_1m_2},
\label{eq:almI}
\eeq
where $I_{ll_1l_2mm_1m_2}$ is defined in
equation (\ref{eq:intyyy}).

\begin{figure*}[t]
\includegraphics[width=3in]{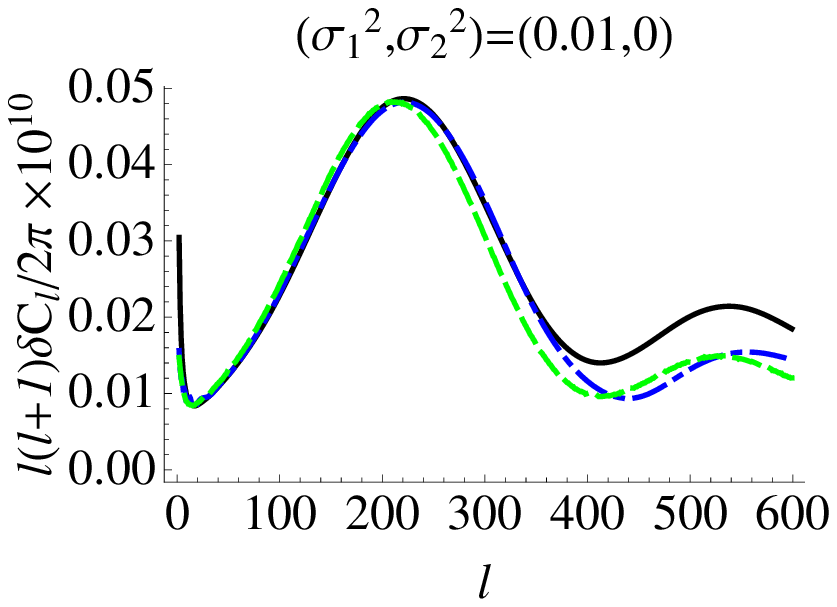}
\includegraphics[width=3in]{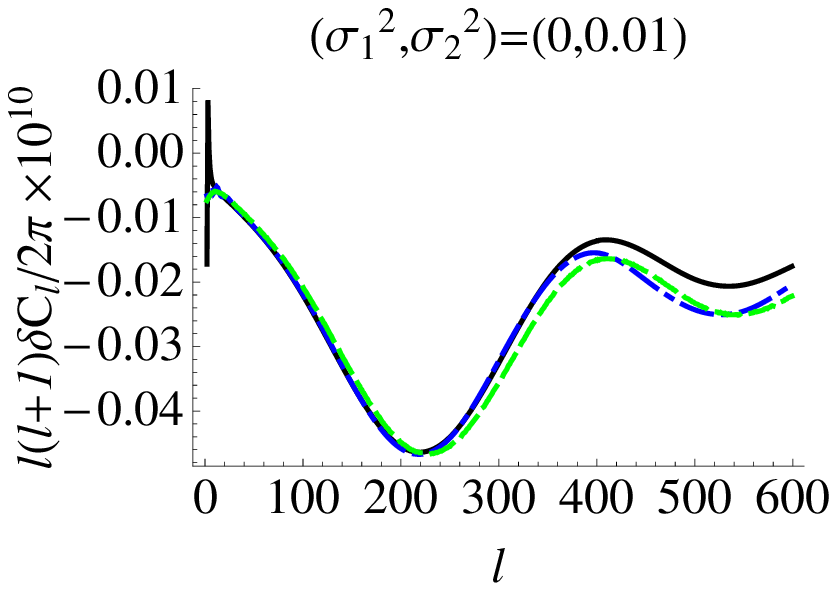}
\includegraphics[width=3in]{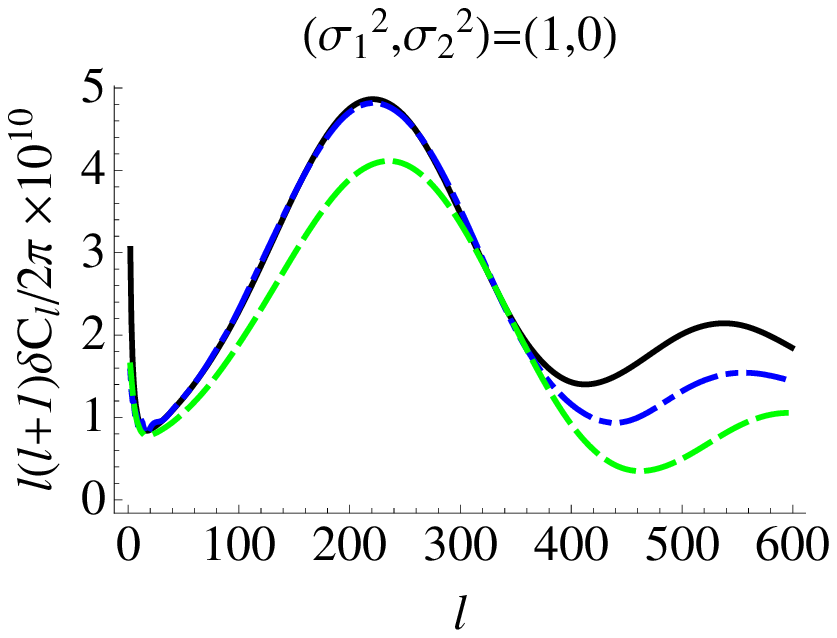}
\includegraphics[width=3in]{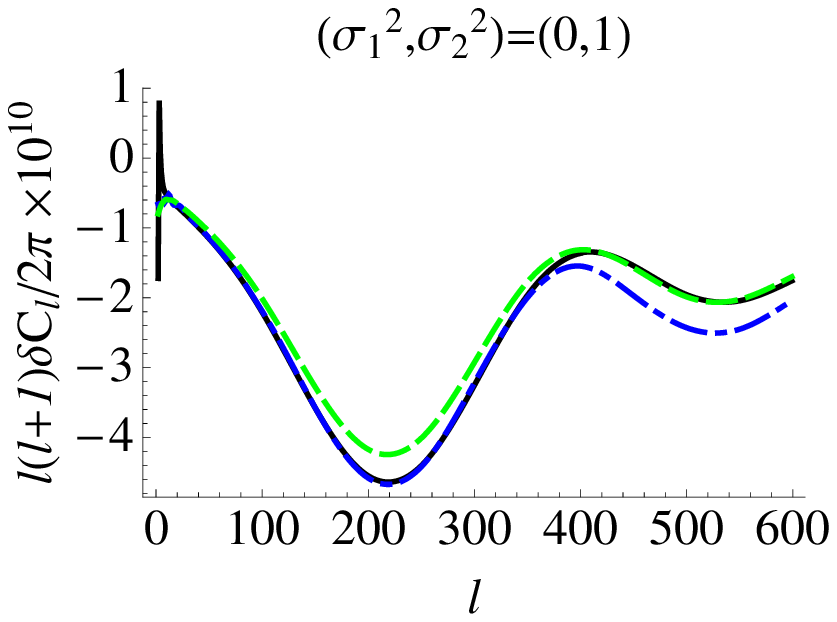}
\caption{Power spectrum changes.
The solid curves are the $\delta C_l$'s resulting
from nonzero $\sigma_1$ or $\sigma_2$ as described in equation
(\ref{eq:dif1}).  The dot-dashed curves are the
results of the linear approximation (\ref{eq:dif2}), with
best-fit $\delta g_i$.  The dashed curves are are the
$\delta C_l$'s resulting from recomputing 
the power spectrum with the modified parameter values.
Power spectra are computed in dimensionless $(\Delta T/T)^2$ form.  In
these units, the CMB power is of order $10^{10}l(l+1)C_l/2\pi\sim 1.5$ - 8
over the multipole range of interest; thus
in the top panels, the deviations in the power spectrum
are all quite small.  The different approximations agree
reasonably well, showing that the approximations
in this section are adequate.
}
\label{fig:ps}
\end{figure*}

In an isotropic
model, the power spectrum is given by $C_l=\langle |a_{lm}|^2\rangle$,
which is independent of $m$.  In an anisotropic
model, this quantity is not necessarily independent of $m$, so
we define the power spectrum to be the average over $m$:
\beq
C_l={1\over 2l+1}\sum_m\langle|a_{lm}|^2\rangle.
\eeq
We substitute equation (\ref{eq:almI}) into this expression.
We then make use of the fact that both the $a_{lm}^{(0)}$ and
$f_{lm}$ coefficients are drawn from isotropic
Gaussian random processes, which implies that 
different coefficients are uncorrelated:
\begin{eqnarray}
\langle a_{lm}^{(0)}a_{l'm'}^{(0)*}\rangle&=&
C_l^{(0)}\delta_{ll'}\delta_{mm'},\\
\langle f_{lm}f_{l'm'}^{*}\rangle&=&
C_l^{(f)}\delta_{ll'}\delta_{mm'},\\
\langle a_{lm}^{(0)}f_{l'm'}^*\rangle&=&0.
\end{eqnarray}
The result is
\begin{eqnarray}
C_l&=&\sum_{l_1,l_2}C_{l_1}^{(0)}C_{l_2}^{(f)}
\left({1\over 2l+1}\sum_{m_1,m_2,m}I_{l_1l_2lm_1m_2m}^2\right)\\
&\equiv&
 \sum_{l_1,l_2}C_{l_1}^{(0)}C_{l_2}^{(f)}\overline{I^{^2}}_{_{\!\!\!\!l_1l_2l}}.
\label{eq:cl}
\end{eqnarray}


The sum inside the parentheses is over all $m$, $m_1$, and
$m_2$ values that make the Wigner 3-$j$ symbols physical.  We assume
that $C_{l_2}^{(f)}=0$ for $l_2 > 2$, so that the double sum above
becomes three single sums:
\begin{eqnarray}
C_l&=&\sum_{l_1}C_{l_1}^{(0)}C_0^{(f)}\overline{I^{^2}}_{_{\!\!\!\!l_10l}}
+
\sum_{l_1}C_{l_1}^{(0)}C_1^{(f)}\overline{I^{^2}}_{_{\!\!\!\!l_11l}}
\nonumber\\
&&\qquad\qquad
+\sum_{l_1}C_{l_1}^{(0)}C_2^{(f)}\overline{I^{^2}}_{_{\!\!\!\!l_12l}}
\end{eqnarray}
Because of the triangle inequality on the 3-$j$ symbols, 
the first sum contains only one nonzero term ($l_1=l$),
and not surprisingly this term reduces to $C_l^{(0)}$.
The second and third sums similarly have only a few nonzero
terms.


Substituting $C_1^{(f)}={\sigma_1^2/2}$ and
$C_2^{(f)}={\sigma_2^2/6}$, we find that the
difference between modulated and unmodulated power spectra is


\begin{equation}
\delta C_l=C_l-C_{l}^{(0)}=
{\sigma_1^2\over 2}\sum_{l_1} C_{l_1}^{(0)}
\overline{I^{^2}}_{_{\!\!\!\!l_11l}}
+
{\sigma_2^2\over 6}
\sum_{l_1=2}^\infty C_{l_1}^{(0)}
\overline{I^{^2}}_{_{\!\!\!\!l_12l}}
\label{eq:dif1}
\end{equation}

We see that  $\delta C_l$ is a linear function of 
$\sigma_1^2$ and $\sigma_2^2$. 
For any given model, we can calculate the $\delta C_l$ contributions
from $\sigma_1^2$ and $\sigma_2^2$ independently.

Assuming that the perturbation from the applied field is small, we expect the
change in the power spectrum, and hence the
change in the inferred parameter values, to be small.
In the standard $\Lambda$CDM paradigm, the observed power
spectrum depends on six parameters:
$\Omega_b$ (baryon density),
$\Omega_{\rm cdm}$ (dark matter density), $\Omega_\Lambda$ (vacuum energy
density), $n$
(spectral index), $H_0$ (Hubble constant, $h={H_0 / (100\,{\rm km\, s^{-1}\,
Mpc})}$), and $A$ (normalization constant for all $C_l$, relative to
the current best fit values from WMAP). 
Calling these parameters $g_1,\ldots,g_6$, 
we have to a good approximation
\begin{equation}
\delta C_l\approx\sum_{i=1}^6 \delta g_i {\partial C_l\over \partial g_i}
\label{eq:dif2}
\end{equation}


Setting equations (\ref{eq:dif1}) and (\ref{eq:dif2}) equal,
and splitting the parameter variations into terms that depend
on $\sigma_1$ and $\sigma_2$, we can write 
\begin{eqnarray}
\sum_{i=1}^6 \delta g_{i,{\sigma_1^2}} {\partial C_l\over \partial g_i}&\approx& 
{\sigma_1^2\over 2}\sum_{l_1=2}^\infty C_{l_1}^{(0)}
\overline{I^{^2}}_{_{\!\!\!\!l_11l}}
\\\sum_{i=1}^6 \delta g_{i,{\sigma_2^2}} {\partial C_l\over \partial g_i}&\approx& 
{\sigma_2^2\over 6}\sum_{l_1=2}^\infty C_{l_1}^{(0)}
\overline{I^{^2}}_{_{\!\!\!\!l_12l}}
\\\delta g_i &=&\delta g_{i,{\sigma_1^2}}+\delta g_{i,{\sigma_2^2}}
\label{eq:linearparam2}
\end{eqnarray}

\begin{figure*}[t]
\begin{tabular}{cccc}
\includegraphics[width=1.5in]{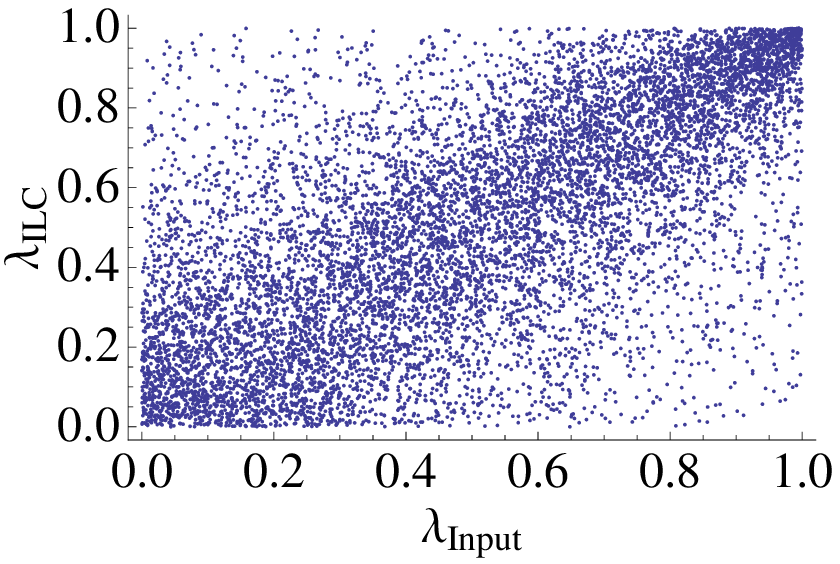}&
\includegraphics[width=1.5in]{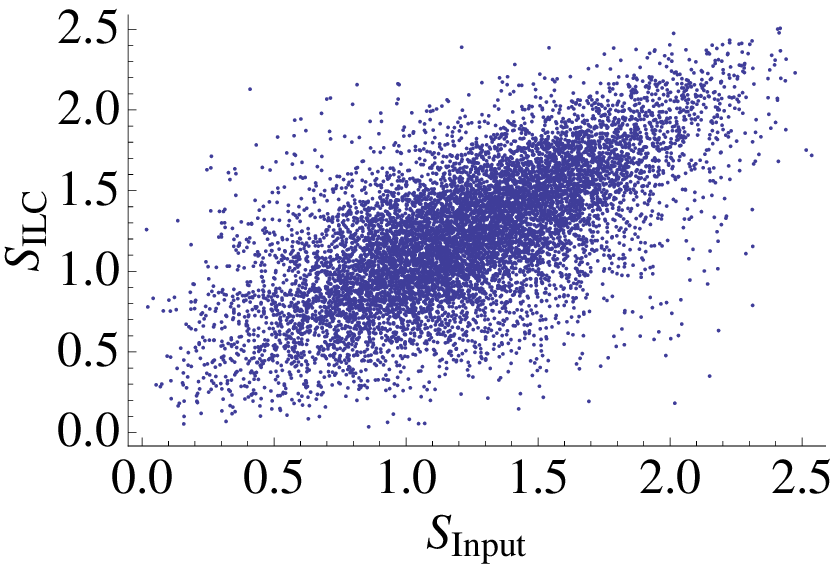}&
\includegraphics[width=1.5in]{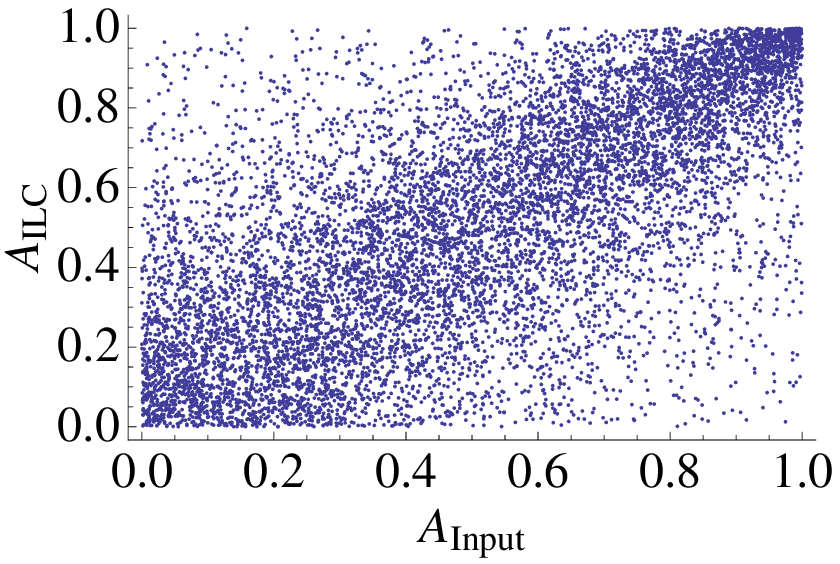}&
\includegraphics[width=1.5in]{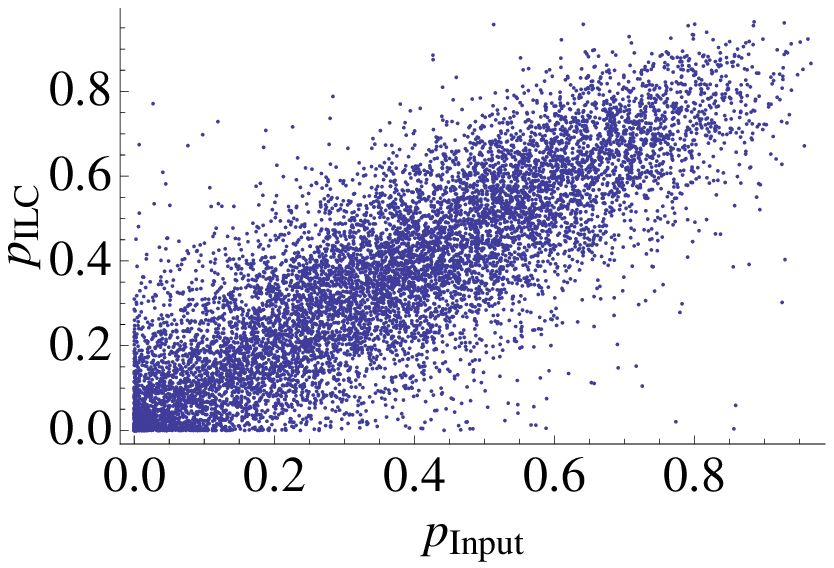}\\

\includegraphics[width=1.5in]{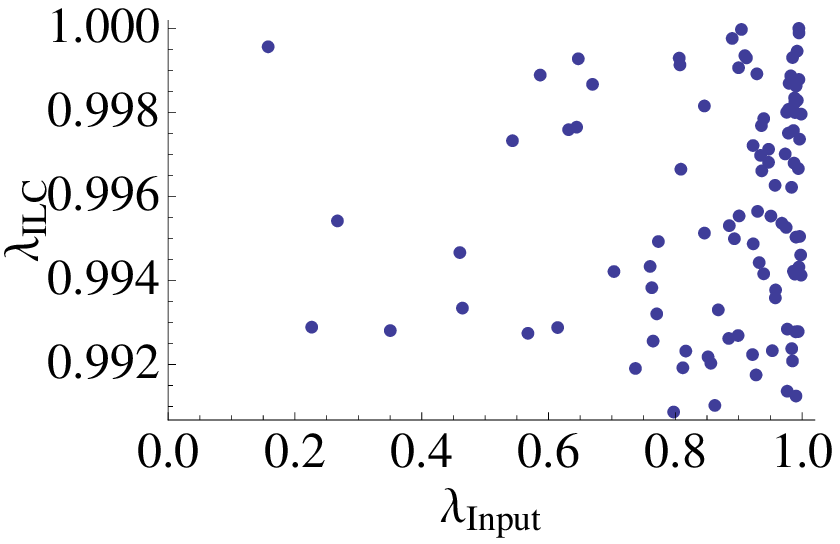}&
\includegraphics[width=1.5in]{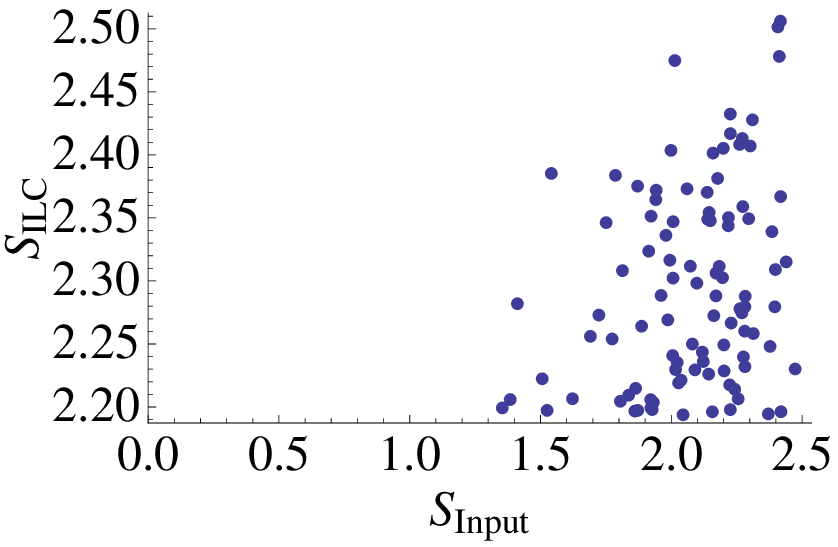}&
\includegraphics[width=1.5in]{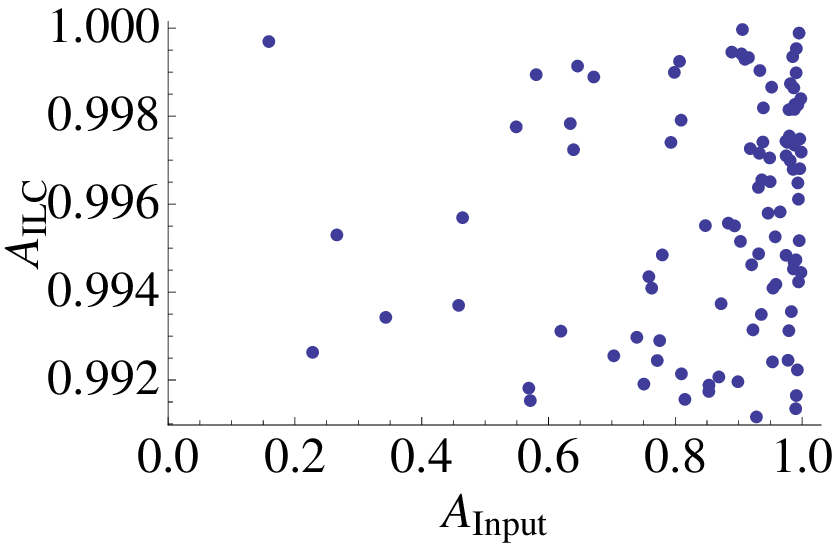}&
\end{tabular}

\caption{
Effect of foreground contamination.  The top row shows scatter plots
indicating the relation between statistics derived
from simulated \cite{eriksenilc05} input (foreground-free) maps and ILC reconstructions that include
residual foreground contamination.  The bottom row shows only those
simulations for which the ILC statistics lie in the top 1\% of 
their distributions.  (The planarity statistic $p$ is not shown
in the bottom row, as it does not yield an anomalously high value
in the actual data.)
}
\label{fig:foregrounds}
\end{figure*}

We use Euler's method to approximate $\partial C_l\over \partial g_i$,
starting from the current best fit values
$\vec{g}^{(0)}=(\Omega_b^{(0)},\Omega_{\rm
cdm}^{(0)},\Omega_\Lambda^{(0)},n^{(0)},h_0^{(0)},A^{(0)})=(0.046,0.224,0.73,0.99,0.72,1)$. We
vary each parameter $g_i$ independently by about 2\% of the
original value, calculate the resulting $C_l$'s with CMBFAST, and obtain
$\partial C_l$ by calculating difference between the new $C_l$'s and
the standard $C_l$'s. We thus obtain $\partial C_l\over \partial g_i$.
Using equation (\ref{eq:dif1}) and starting with the standard-model
parameter values, we compute the $\sigma_1^2$
and $\sigma_2^2$ contributions to $\delta C_l$.

We can then find best-fit values of $\delta g_{i,\sigma_j^2}$.
We perform a least-squares fit over the range $2\le l\le 600$, 
with weights given by the combination of cosmic variance
and noise errors for WMAP.
To test the validity of this procedure, we compute a
new set of $C_l$'s using CMBFAST with parameters given by 
$\vec g^{(0)}+\delta\vec g$.
Figure \ref{fig:ps} shows that the fitting
works very well, and that the linearity of the $C_l$'s in the specific
direction of $\delta\vec{g}$ validates the approximation 
in equation
(\ref{eq:dif2}). 
For $\sigma_1,\sigma_2$ of order 1, linearity starts to break down,
but such large values
are probably unphysical in any case.

Numerically, in the linear regime the changes in parameters can be calculated
by
\begin{equation}
\delta \vec{g}=
\begin{pmatrix}
\delta\Omega_b\\\delta\Omega_{\rm cdm}\\\delta\Omega_\Lambda\\\delta n\\\delta h\\\delta A
\end{pmatrix}\approx
\begin{pmatrix}
3.86\times 10^{-4}&0.963\times 10^{-4}
\\-6.33\times 10^{-3}&-5.78\times 10^{-3}
\\5.24\times 10^{-3}&4.26\times 10^{-3}
\\-8.39\times 10^{-3}&-7.88\times 10^{-3}
\\4.74\times 10^{-3}&6.55\times 10^{-3}
\\-0.285&-0.753
\end{pmatrix}\begin{pmatrix}
\sigma_1^2\\\sigma_2^2
\end{pmatrix}
\end{equation}

In all cases except for the overall normalization $A$, the parameter
changes are small even for relatively large $\sigma_1^2,\sigma_2^2\sim 1$.
Moreover, as can be seen in Figure \ref{fig:ps}, the residuals $\delta C_l$
have similar shape to the input power spectrum $C_l$ (although
with a negative prefactor for $\sigma_2$), indicating
that the chief error in the linear approximations in this 
section applies to the normalization.  We conclude that, in 
a model of the form considered here, one should take care to recompute
the overall normalization, which affects the normalization
of the matter power spectrum, but that other parameters are likely
to remain approximately unchanged.

\section{Foregrounds}
\label{sec:foregrounds}

The significance of the observed anomalies depends on the choice of data
set (e.g., \cite{landmag}).  We chose to work in spherical harmonic space,
leading to the requirement of an all-sky data set.  We thus worked with
the WMAP ILC data.  With this choice
of data set, one must wonder about the effect of 
residual foreground contamination on our results.

We can begin to assess these effects using a set of 10\,000 publicly available
ILC simulations \cite{eriksenilc05}.  For each simulation, both
the foreground-free input map and the ILC reconstructions, with
residual foreground contamination, are provided.  In each case, we computed the 
four statistics discussed in this paper, namely the angular-momentum
statistic $\lambda$, the Schwarz et al. multipole vector statistic $S$, 
the multipole vector alignment statistic $A$, and the planarity statistic
$p$.  Figure \ref{fig:foregrounds} shows a comparison of the input and 
and ILC maps for each statistic.  In each case, there is a strong correlation,
but the scatter is considerable.

The probability density for each of the four statistics undergoes no
significant change between the input and ILC ensembles.  (In other words,
in each of the four plots on the top row of Figure \ref{fig:foregrounds},
histograms of the $x$ and $y$ values look essentially identical.)  This
can be quantified in a variety of ways.  Since
we are most interested in the probability distribution near the upper
tail of the distribution of each statistic (except for $p$, which
has negligible effect on any of our conclusions anyway), we look at the behavior
of the distributions near the 99th percentile.  For each statistic, we 
find the 99th-percentile value in the 10\,000 ILC maps, and count the
number of input maps lying above that value.  (The results are essentially
identical if the two roles are reversed.)  If the input and ILC 
probability densities are 
the same, we expect to find 100 in each case.  The actual values found
deviate from this expected value by $1,1,5,-6$ 
for $\lambda,S,A,p$ respectively.  All are well within the $10\%$ fluctuation
level
expected due to Poisson noise.

From this test, we can conclude that foregrounds do not significantly alter
the statistical significance of anomalies based on these statistics.
Due to the problem of a posteriori statistics, reasonable people
can disagree about whether to take the ILC multipole alignment seriously,
but one's opinion on this question need not be altered by 
consideration of foreground contamination.

In this paper we do not chiefly address the question of whether
the multipole alignment is statistically significant; on the contrary,
we provisionally
adopt the stance that it is and ask what form an explanation of it might
take. For this sort of question, we need to go beyond the simple considerations
above and consider the correlations between input and ILC maps.  After all,
nonstandard cosmological models such as the broken-isotropy models we consider
affect the probability of seeing multipole alignments in the foreground-free
(``input'') maps, whereas the likelihoods that form the basis of our evidence
calculations are based on the ILC map.  

Once again, 
for the three statistics $\lambda,S,A$ that primarily affect our results, 
we are interested in the relation between input and ILC values
near the upper end of the statistics' ranges.  Specifically, we want to know
whether the observed large ILC value implies a large input 
value in the foreground-free CMB.  The bottom row of plots in Figure
\ref{fig:foregrounds} provides one qualitative way of addressing this question.
For each statistic, we show a scatter plot comparing input and ILC values
as in the upper row, but showing only points corresponding
to the top 1\% of ILC values.  Many points cluster near the right,
indicating that a high ILC value is likely, but by no
means certain, to have come from a high
input value.

Let us be slightly more quantitative.  For any given
statistic, say $\lambda$, 
we extract the realizations
for which the ILC maps are anomalously high, lying in the top 1\%
of the distribution.  For these 100 realizations, we find
the value of the statistic in the input map, $\lambda_{\rm Input}^*$,
and look at its ranking in the full set of 10\,000 input realizations.
This gives
the cumulative probability $P_{\rm input}\equiv {\rm Pr}[\lambda_{\rm Input}<
\lambda_{\rm Input}^*]$ for each of the 100 input maps.  If foreground
contamination were negligible, then these 100 maps would
lie in the top percentile of the input distribution, i.e., all 100
$P_{\rm input}$ values would be above 0.99.

\begin{figure}
\includegraphics[width=3in]{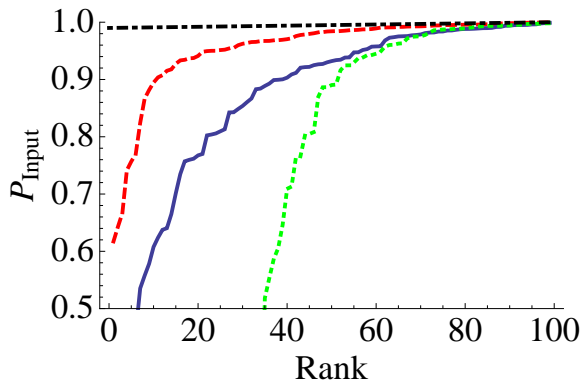}
\caption{
For the three statistics $\lambda$ (solid), $S$ (dashed), $A$ (dotted),
we select the top 1\% of ILC simulations, and determine the cumulative
probability $P_{\rm Input}$ 
of the statistic in the input map.  The values are sorted and plotted.
In the absence of any effect from foregrounds, ILC and input
maps would be identical, and the result would be a 
straight line extending from 0.99 to 1 (dot-dashed line).
}
\label{fig:foregrounds2}
\end{figure}

Figure \ref{fig:foregrounds2} shows the result of this exercise
for each of the three statistics $\lambda,S,A$.  In each case,
the results are sorted by the value of the statistic in the input data.
The results show that the statistic $S$ is least affected by foreground
contamination: if a realization lies in the top 1\% of the ILC maps, 
there is a high probability that it also lies near the top of the probability distribution
of the input maps as well.  For the three statistics $\lambda,S,A$, the
median values of $P_{\rm Input}$ for the ILC top 1\% maps are 93.2\%, 98.5\%,
89.0\%, as compared to the value 99.5\% that would occur
if there were no foregrounds.

Generically, if the correlation between input and ILC maps is weak,
then we would expect the enhanced likelihood and Bayesian evidence results
of Section \ref{sec:align} to be overestimates of the correct results.
Intuitively, this seems clear: if the connection between the true CMB and
the observed ILC data is weak, then so is our ability to draw cosmological
conclusions from the ILC data.  We can express this idea more formally
as follows.  Our theoretical models allow us to calculate probability
distributions for the ``input'' data (i.e., the pure CMB), while
our observations are of the ILC data.  The correct procedure, therefore,
is to convert the input probability distributions into ILC
probability distributions by convolution with a conditional probability
function $P({\rm ILC}|{\rm Input})$.  Such a convolution would 
smooth out variations in likelihood.

We conclude, therefore, that because of foreground contamination, 
the results shown in Section \ref{sec:align} should be regarded as upper limits.
The effect of foregrounds on the results is difficult to quantify, but
based on Figure \ref{fig:foregrounds2} we expect it to be smallest for
the results based on the Schwarz et al.\ statistic $S$.

\section{Discussion}
\label{sec:discussion}

The various anomalies that have been noted in the large-angle CMB
may provide hints of departures from the standard cosmological
model, possibly including violations of statistical isotropy.
Although the statistical significance of these anomalies is
difficult or even impossible to quantify a posteriori, these possibilities
are exciting enough to warrant closer examination.

We have considered several classes of physically-motivated models
that might explain the anomalies.  We have calculated Bayesian evidence
ratios to assess the degree to which the purported anomalies in the
multipoles $l=2,3$ favor
the anisotropic models over the standard model.

According to the pioneering work of 
Jeffreys \cite{jeffreys}, a Bayesian evidence ratio
constitutes ``substantial'' evidence if $\ln\Lambda>1$ and ``strong''
evidence if $\ln\Lambda>2.5$.  As the results in the Section \ref{sec:align}
make clear (note that what is plotted in each case is $\Lambda$, not
$\ln\Lambda$), only for the most judicious choice of prior do the 
tests performed here reach the ``substantial'' level, and they never 
come close to being ``strong.''

Of course, Jeffreys's criteria are somewhat arbitrary, but in this case 
they seem to describe the situation fairly well.  Recall that the
evidence ratio $\Lambda$ is simply the factor by which the ratio
of prior probabilities must be adjusted, in the light of the observations,
in order to get the posterior probability ratio.  Presumably, the prior
probability distribution assigns very low weight to the less natural
anisotropic models, so even after applying an evidence ratio $\Lambda\sim 3$,
the anisotropic models are still considered unlikely.  
One would require
an exponentially large evidence ratio before assigning significant probability
to the anisotropic models.

We used several different statistical approaches to to 
characterize the observed multipole alignment.  Some ($\lambda,S$)
are adopted from previous work, while others $(A,p)$ are of our own devising.
In the latter case, we attempted to minimize (although not eliminate)
the problem of a posteriori statistics by choosing a method blindly
that seemed to us to naturally encapsulate the observed phenomena with
minimal arbitrary choices.  In any case, the general consistency of 
the results based on the different statistics indicates that the
approach we have followed is robust.

We have estimated the changes in cosmological parameter estimates 
that would arise if the anisotropic models were shown to be correct.
The chief effect of the modulation is on the estimate of the overall
power spectrum normalization, which would of course have consequences
for studies of large-scale structure.
Our calculations are valid only if the modulation is applied to
the CMB at all $l$-values measured by WMAP.  If
a more complicated
model is correct (e.g. \cite{hou}), in which only some scales are modulated,
then the parameter changes would presumably be smaller.

We have used simulations of the ILC mapmaking process to evaluate the
degree to which foreground contamination might affect our results.
The statistic $S$ appears least affected by this problem: ILC maps with
high values of $S$ are very likely to correspond to high values of $S$ in
the intrinsic CMB.  A thorough treatment of foregrounds in our analysis
would generically reduce the (already modest) enhancements in the evidence
ratio, so due to the effects of foregrounds 
our results can be regarded as upper limits.

In this paper, we have tentatively adopted the point of view that there
are anomalies to be explained.  Of course, one would greatly prefer to
settle this question in a way that was not plagued by the problem
of a posteriori statistics.  To do this, we would require a new data
set that probes similar scales to the large-angle CMB.    
All-sky polarization maps may provide
some insight into these issues \cite{frommert,dvorkin}.  Another
possibility is to survey the ``remote quadrupole'' signal
found in the polarization of CMB photons scattered in 
distant clusters \cite{kamloeb}, which can be used to reconstruct information
on gigaparsec-scale perturbations \cite{bunnrq,abramorq}.
Although
gathering data on these scales 
is a difficult task, the potential for learning about the
structure of the Universe on the largest observable scales makes
it worth pursuing.

\section*{Acknowledgments}

We thank an anonymous referee for very helpful comments.  EFB is
grateful for the hospitality of the Laboratoire APC at the Universit\'e Paris
VII, where some of this work was performed.  This work was supported by NSF
awards  0507395 and 0922748.

\bibliography{anisotropic}

\end{document}